\RequirePackage{fix-cm}
\documentclass{svjour3}                     
\smartqed  
\usepackage{graphicx}
\usepackage{multirow}
\usepackage[T1]{fontenc}
\usepackage[utf8]{inputenc}
\usepackage{array}
\usepackage{longtable}
\usepackage{rotating}
\usepackage{lscape}

\newcommand{\be}{\begin{equation}}
\newcommand{\ee}{\end{equation}}
\def\dif{\mathop{\rm \,d}\nolimits}

\begin{document}
%
%
\title{Relativistic perfect fluids in local thermal equilibrium}
\titlerunning{Relativistic perfect fluids in local thermal equilibrium}        

\author{Bartolom\'e Coll, Joan Josep Ferrando        \and
   Juan Antonio S\'aez 
}


\institute{Bartolom\'e Coll \at
              Departament d'Astronomia i Astrof\'{\i}sica, Universitat
de Val\`encia, \\ E-46100 Burjassot, Val\`encia, Spain. \and
Joan Josep Ferrando \at
              Departament d'Astronomia i Astrof\'{\i}sica, Universitat
de Val\`encia, \\ E-46100 Burjassot, Val\`encia, Spain. \\
Observatori Astron\`omic, Universitat
de Val\`encia, \\ E-46980 Paterna, Val\`encia, Spain. \\
                           \email{joan.ferrando@uv.es}           
           \and
          Juan Antonio S\'aez  \at
               Departament de Matem\`atiques per a l'Economia i l'Empresa,
Universitat de Val\`encia, \\E-46022 Val\`encia, Spain
}

\date{Received: date / Accepted: date}

\maketitle

\begin{abstract}
Every evolution of a fluid is uniquely described by an energy tensor. But the converse is not true: an energy tensor may describe the evolution of different fluids. The problem of determining them is called here the {\em inverse problem}. This problem may admit unphysical or non-deterministic solutions. This paper is devoted to solve the inverse problem for perfect energy tensors in the class of perfect fluids evolving in local thermal equilibrium (l.t.e.). The starting point is a previous result (Coll and Ferrando in J Math Phys 30: 2918-2922, 1989) showing that thermodynamic fluids evolving in l.t.e. admit a purely hydrodynamic characterization. This characterization allows solving this inverse problem in a very compact form. The paradigmatic case of perfect energy tensors representing the evolution of ideal gases is studied in detail and some applications and examples are outlined.

\end{abstract}

\section{Introduction. The inverse problem for perfect energy tensors}
  A {\em perfect fluid} $f$ means here a Pascalian\footnote{A fluid is said Pascalian if it has zero viscosity and, for an observer at rest, its
stress tensor is isotropic.} fluid of zero heat conductivity. 
In a domain $\Omega$ of the space-time,%
\footnote{In the present context, it does not matter whether
the fluid contributes to the gravitational field or is a test
fluid in any given space-time.} %
to every one of its possible evolutions corresponds an {\em
energy tensor}%
\footnote{We follow here the extended use of calling {\em energy tensor} the field of energy tensors at every point of the space-time domain.} %
$T_{f}$ of the form   
\begin{equation}
\label{perfect-energy}
T_{f} = (\rho_{f} + p_{f}) u_{f}\otimes u_{f}
+ p_{f} \, g \, ,
\end{equation}
where $g$ is the space-time metric and $\rho_{f},$ $p_{f}$ and $u_{f}$ are respectively the {\em energy density}, {\em
pressure} and {\em unit velocity} of the fluid. In the absence of exterior constraints, this energy tensor is divergence-free:
\begin{equation} \label{divergence-free}
 \nabla \cdot T_{f} = 0 \, .
\end{equation}
Tensor fields of the form (\ref{perfect-energy}) and satisfying (\ref{divergence-free})
will be called {\em perfect energy tensors} herein.

\begin{figure}[h]
{\vspace{-0mm}\includegraphics[width=0.5\textwidth]{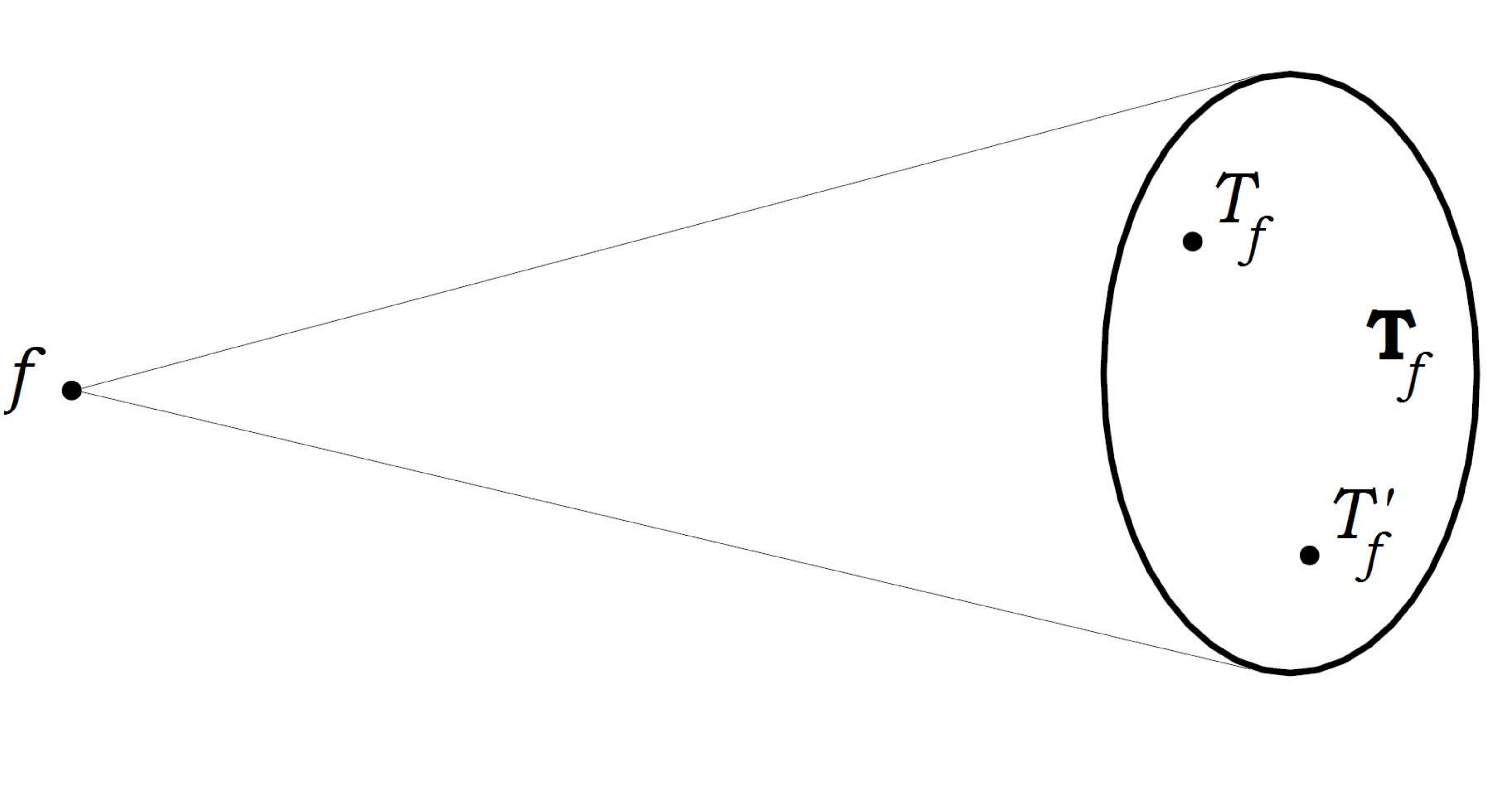}} \quad   \quad {\includegraphics[width=0.45\textwidth]{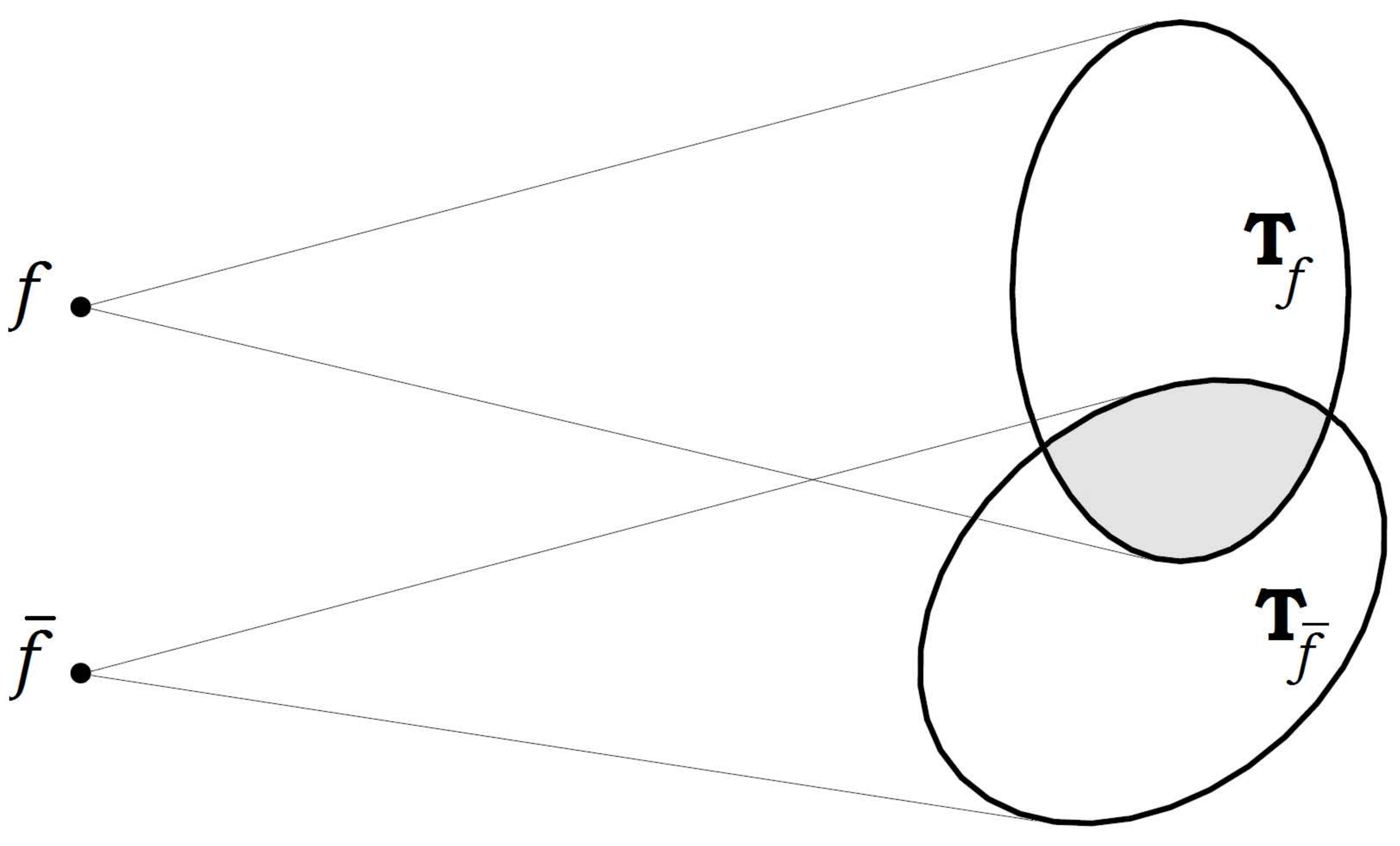}}
\caption{On the left, the set ${\bf
T}_{f}$ of perfect energy tensors corresponding to all the possible evolutions of a fluid $f$. On the right, two non disjoint sets ${\bf T}_{f}$ and ${\bf
T}_{{\bar f}}$ corresponding to two different fluids $f$ and $\bar f$. \label{Fig-1-2}}
\end{figure}

Thus, {\it a set} ${\bf T}_{f} \equiv \{T_{f}\} $ is associated to {\em every} perfect fluid $ f $, namely those perfect energy tensors corresponding to {\em every one of} its possible evolutions in the space-time domain $\Omega$. The {\em energetic description} of every evolution of the perfect
fluid $f$ consists of the specification of the perfect energy
tensor $T_{f}$ that corresponds to this 
evolution, i.e. to the specification of the distribution in $\Omega$ of its
{\em hydrodynamic variables} $\rho_{f},$  $p_{f}$ and $u_{f}.$  The left-hand diagram in
Figure \ref{Fig-1-2} outlines this situation.

It becomes obvious that the sole energetic description of a perfect 
fluid is insufficient to characterize it physically. In
other words, the sets ${\bf T}_{f}$ and ${\bf T}_{{\bar f}}$ of perfect energy tensors corresponding to all possible evolutions of two different perfect fluids $f$ and $\bar f$ are not necessarily disjoint, ${\bf T}_{f} \cap {\bf
T}_{{\bar f}} \not= \emptyset,$ so that the equation $T_f = T_{\bar f}$ for different perfect fluids $f$ and $\bar f$
may admit non vanishing solutions $T$. The right-hand diagram in Figure \ref{Fig-1-2} illustrates this situation.

\begin{figure}
{\includegraphics[width=0.45\textwidth]{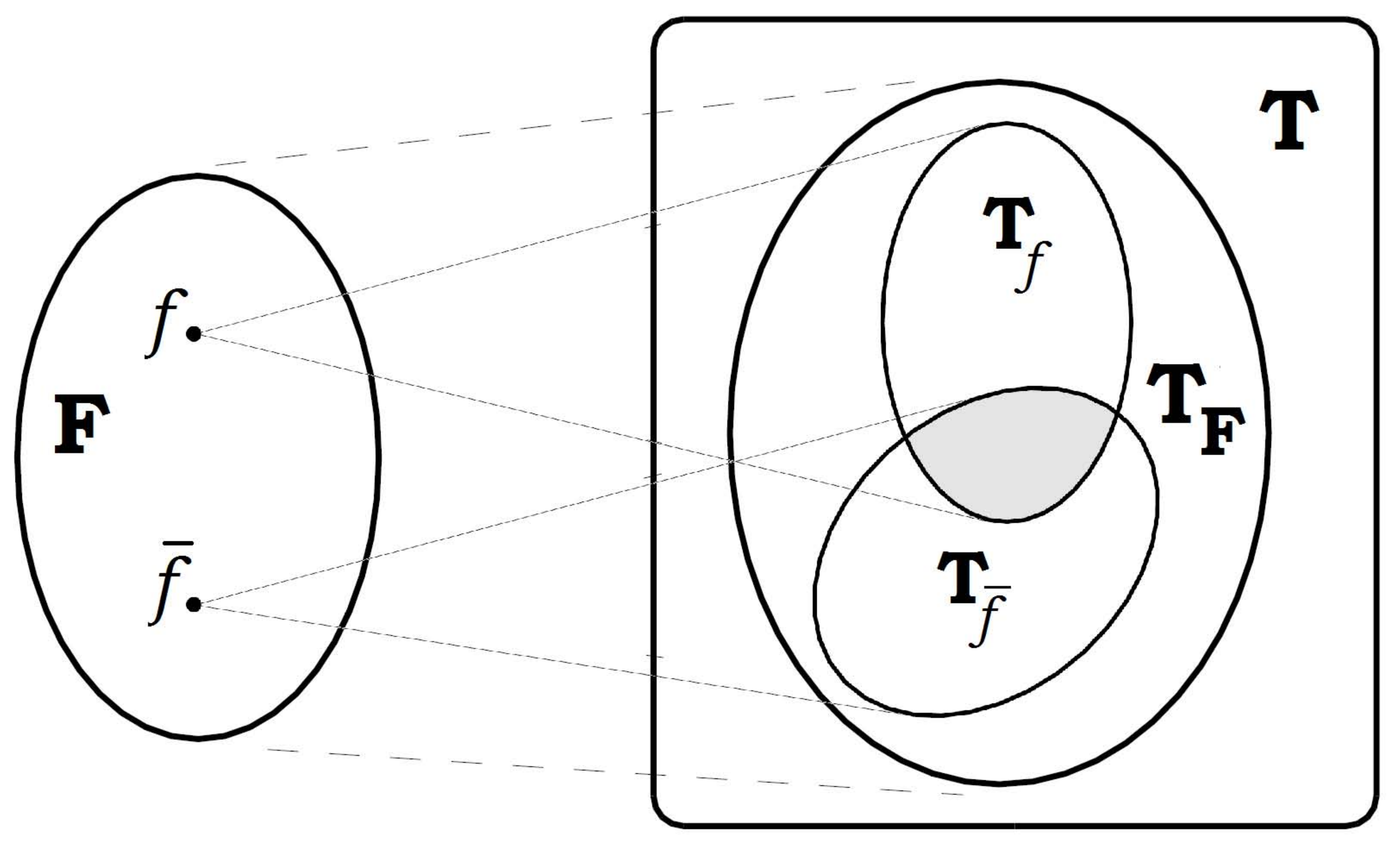}} \ \quad \quad \quad  {\includegraphics[width=0.455\textwidth]{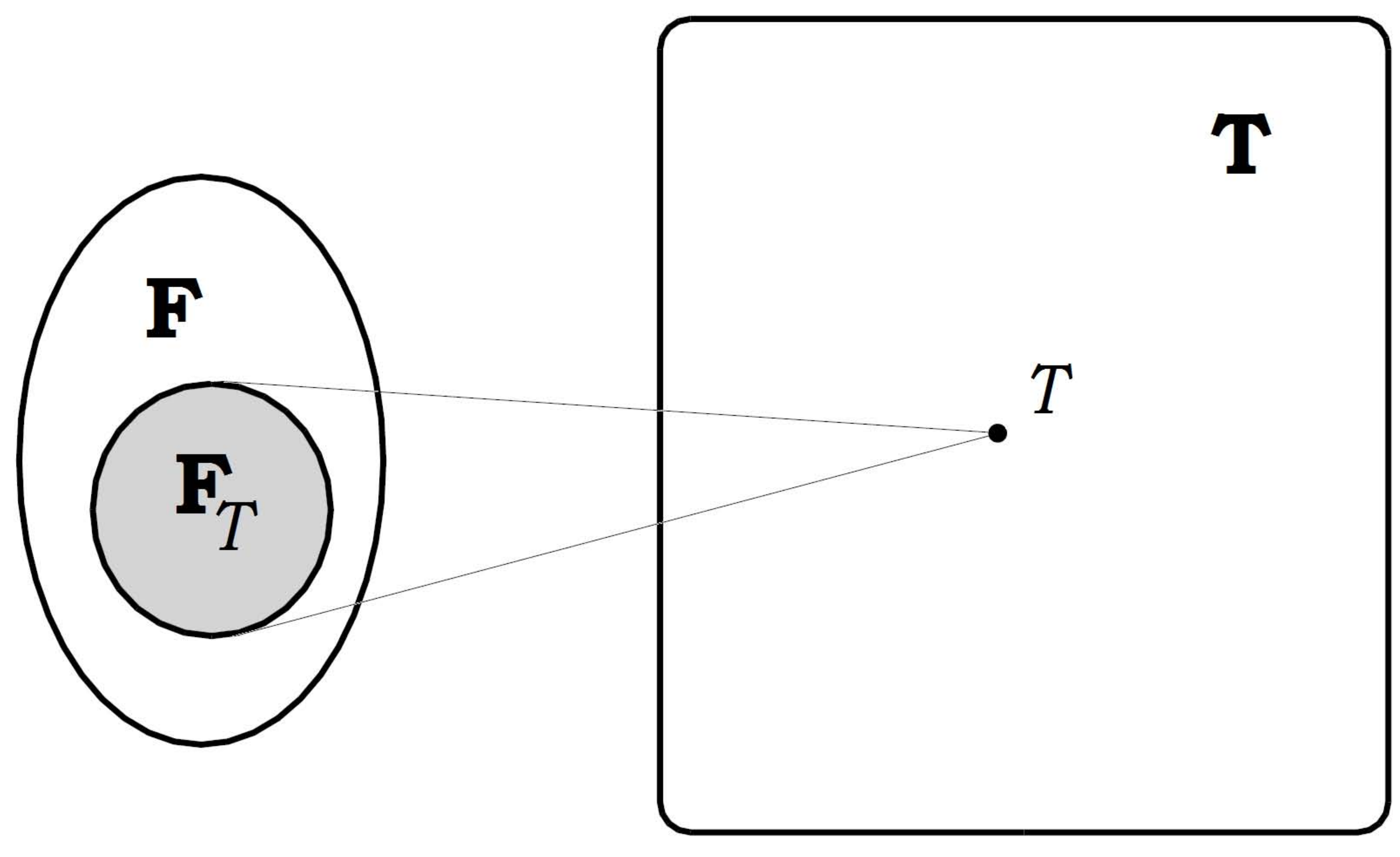}} 
\caption{The left-hand diagram shows the set ${\bf T}_{\bf F}$ of all perfect energy tensors $T_f$, corresponding to all possible
evolutions of all perfect fluids, as a strict subset of the set ${\bf T}$ of all perfect energy tensors. On the right, a perfect energy tensor $T$ corresponds to the evolution of a whole class ${\bf F}_{T }$ of fluids. \label{Fig-3-4}}
\end{figure}

On the other hand, perfect energy tensors may be generated without any relation to perfect fluids.%
\footnote{Perfect energy tensors unrelated to perfect fluids may be generated both theoretically and experimentally. Theoretically, it is
sufficient to consider arbitrary solutions $\rho,$ $p,$ $u$ to the divergence free condition. Experimentally it is sufficient
to take into account particular {\em perfect evolutions} (e.g. static, isothermal, homogeneously strained) of otherwise generically {\em
non perfect}, anisotropic fluids.} %
 This means that, if ${\bf T}$ denotes the set of all perfect
energy tensors $T$, $\, {\bf T} = \{T \mid T\! =\! (\rho\! +\! p)u\otimes u + p\,g ,\  \nabla \cdot T\! =\! 0 \}$, if ${\bf F}$
denotes the set of all perfect fluids $f,$ ${\bf F} = \{f\},$ and if ${\bf T}_{\bf F}$ denotes the set of all
perfect energy tensors $T_f$ corresponding to all possible evolutions of all perfect fluids, ${\bf T}_{\bf F} = \{{\bf T}_f,
\forall  f \in {\bf F}\},$ then ${\bf T}_{\bf F}$ is strictly contained in ${\bf T}:$ ${\bf T}_{\bf F} \subset {\bf T}$. 
In other words, there exist perfect energy tensors $T$  in ${\bf T}$  that {\em do not correspond} to (any particular
evolutions of) perfect fluids. The left-hand diagram in Figure \ref{Fig-3-4} illustrates this situation.

But when the perfect energy tensor $T$ may correspond to a perfect fluid, it will generically correspond,
not to a sole perfect fluid $f$, but to a whole class ${\bf F}_{T }$ in ${\bf F},$ as shown in the right-hand diagram in Figure \ref{Fig-3-4}.

Generically, the fluids considered in physics are  {\em local} (i.e. their configuration at every event depends only on its variation in the neighborhood of the event) and {\em deterministic} (i.e. their evolutions are  univocally determined by their past). Consequently, their evolutions  have to be described by a {\em closed} system of {\em differential equations} (i.e. a system admitting unicity of the Cauchy problem). Following this usage, from now on all the perfect fluids considered herein will be local and deterministic although, for the sake of brevity, we shall not mention it. {\em A fortiori}, all the above relations among perfect fluid evolutions and perfect fluid energy tensors remain true for local and deterministic perfect fluids. 

Usually, in many physical situations one starts from a known perfect fluid $f$ of ${\bf F}$ and looks for energetic descriptions $T_{f}$ in  ${\bf T}_{f}$ corresponding to some particular evolutions of $f$; here we are interested in the other way round:\\[2mm]
{\bf Inverse problem}. {\em The {\em inverse problem for a
perfect energy tensor} $T$, is that of the existence and determination of the
non empty set ${\bf F}_{T}$ of all perfect fluids $f$ for which $T$ is the energetic description of a particular evolution}. \\ 

The main aim of the present paper is to give a precise meaning to this inverse problem and to solve it.

As described above, the set ${\bf T}_f$ of perfect energy tensors associated  to an arbitrary perfect fluid $f$ describes the energetic characterization of all the evolutions of $f$. And as $f$ is a local and deterministic perfect fluid, the set of its evolutions amounts the set of the different {\em initial configurations} that it  may adopt. 

There are also perfect energy tensors in ${\bf T}$ that do not correspond to the evolution of any initial conditions of any perfect fluid. Consequently, it becomes {\em meaningless} to try to find the physical interpretation of an energy tensor $T$ submitted to the sole divergence-free condition. Indeed, there are essential ambiguities inherent to the macroscopic thermodynamic models concerning the lack of sufficient conditions to be imposed on a model of fluid in order to represent a realistic "physical fluid". 

Commonly accepted necessary conditions of physical reality for arbitrary macroscopic media are the {\em energy conditions},\footnote{First stated by Pleba\'nski \cite{Plebanski}, and also considered by Hawking and Ellis in \cite{Hawking-Ellis}, who seemed unaware of Pleba\'nski work.} but, if they may be considered as (algebraic) necessary conditions, they are manifestly insufficient to select
realistic physical fluids.

Indeed, the divergence-free condition is {\em incomplete} from a deterministic point of view, so that to every initial configuration corresponds, not a sole perfect energy tensor, but a partially arbitrary family of them. In order to
select a physical (deterministic) evolution, one has to complete the above divergence-free condition suitably, i.e. to find a {\em deterministic closure} for it.%
\footnote{To choose a {\em deterministic closure} for the energy conservation equations means to complete these equations with some other physically meaningful ones so as to obtain a closed system in the sense considered above, i.e. a system admitting a unique evolution for every initial configuration.} %
 All the  deterministic closures proposed in the literature for arbitrary (non barotropic\footnote{For the notions of {\em barotropic evolution}, {\em barotropic energy tensor} and {\em barotropic perfect fluid}, see below.}) perfect fluids are thermodynamic closures,%
\footnote{These thermodynamic closures have been obtained in relativity from many different approaches, and with very different results. They essentially started with Eckart's third paper on thermodynamics of irreversible processes,
its later contrast with Landau and Lifchitz's point of view gave rise to a great number of new propositions, until the works by Israel, Steward and Marle, or the more recent relativistic version of extended thermodynamics.} 
as they involve new {\em thermodynamic quantities}. Thus, besides the hydrodynamic quantities $(u, \rho, p)$ the hypothesis of {\em local thermal equilibrium}\footnote{see Sect. \ref{sec-lte}.} (l.t.e.) implies introducing at least the {\em matter density} $r$, the {\em specific internal energy} $\epsilon$, the {\em temperature} $\Theta$, and the {\em specific entropy} $s$.

Some years ago \cite{Coll-Ferrando-termo} we showed that by adding to the divergence-free system one suitable equation on the hydrodynamic variables $(u, \rho, p)$ one obtains an equivalent formulation of the l.t.e. scheme. This result allowed to solve the generic  {\em direct problem} for any perfect fluid $f$, namely the determination of the set ${\bf T}_f$ of perfect energy tensors $T_f$ corresponding to all its possible evolutions in l.t.e. Indeed, we determined the necessary and sufficient conditions for the set ${\bf T}_{\bf F}$  to be that of all the perfect energy tensors  corresponding to all the possible evolutions in l.t.e. of all perfect fluids ${\bf F}$.

In Sect. \ref{sec-lte} we introduce the basic variables and relations that usually define a l.t.e. scheme and summarize its above-mentioned hydrodynamic characterization.

Sect. \ref{sec-invers-problem} is devoted to solving the generic inverse problem: to determine, for a perfect energy tensor $T$, the set ${\bf F}_{T}$ of all perfect fluids for which $T$ is the energetic description of a particular evolution. We also analyze the degenerate cases of barotropic or/and isobaroenergetic\footnote{See below for the definition of this notion.} evolutions separately.

The direct and inverse problems stated above may be restricted to a specific family of fluids ${\bf G} \subset {\bf F}$. We can look for the set ${\bf T}_{\bf G}$ of perfect energy tensors $T$ corresponding to particular evolutions of the fluids of ${\bf G}$ (specific direct problem). And given a $T \in {\bf T}_{\bf G}$, we can look for each associated thermodynamic scheme that defines a fluid $f$ of ${\bf G}$ (specific inverse problem). In Sect. \ref{sec-ideal-gas} we study these direct and inverse restricted problems when ${\bf G}$ is the set of generic ideal gases.

Sect. \ref{sec-remarks} is devoted to pointing out the intrinsic, deductive, explicit and algorithmic character of our results, and to remark their interest in building a Rainich-like theory for the family of solutions of Einstein equations corresponding to a specific physical fluid. We also comment the interest of our results in studying the physical reality of known solutions of Einstein equations, and we offer an algorithm in four steps for detecting the solutions that can be interpreted as an ideal gas.


\section{Local thermal equilibrium: a hydrodynamic characterization}
\label{sec-lte}

The energetic description of the evolution of a perfect fluid is given by its perfect energy tensor $T$ in the hydrodynamic variables $\rho$, $p$ and $u$:
\begin{equation}
\label{perfect-energy-tensor}
T = (\rho+ p) u \otimes u + p \, g \, .
\end{equation}
The divergence-free condition for $T$, $\nabla \cdot T = 0$, leads to the conservation equations:
\begin{equation}
d p  + \dot{p} u + (\rho + p) a = 0 \, ,  \label{con-eq1}
\end{equation}
\begin{equation}
\dot{\rho} + (\rho+ p) \theta = 0 \, ,   \label{con-eq2}
\end{equation}
where $a$ and $\theta$ are, respectively, the acceleration and the
expansion of $u$, and where a dot denotes the directional derivative, with respect to $u$, of a quantity $q$, $\dot{q} = u(q) = u^{\alpha} \partial_{\alpha} q$.

A {\em barotropic evolution} is an evolution along which the {\em barotropic relation} $d\rho \wedge dp = 0$ is fulfilled. A perfect energy tensor describing energetically a barotropic evolution is called a {\em barotropic perfect energy tensor}. 

An evolution is said {\em isoenergetic}\footnote{Because l.t.e. concerns only thermal equilibrium of every volume element, the isoenergetic condition is defined here as an evolution property, $\dot{\rho}=0$, not as a spatial one.} if $\dot{\rho}  = 0$, and {\em isobaric} if $\dot{p}  = 0$. For short, an evolution isoenergetic and isobaric is here called {\em isobaroenergetic}.

The energy density $\rho$ may be decomposed in terms of the {\em matter density} $r$ and the {\em specific internal energy} $\epsilon$ in the form: 
\begin{equation} \label{ro-r-epsilon}
\rho= r(1+\epsilon)  \label{masa-energia} \, ,
\end{equation}
requiring the conservation of matter:
\begin{equation}
\nabla \cdot (ru) = \dot{r} + r \theta = 0 \, .  \label{c-masa}
\end{equation}
Then, according to a classical argument, it is always possible to identify an integral divisor of the one-form $\Lambda \equiv d \epsilon + p d (1/r)$  with the (absolute) {\em temperature} $\Theta$ of the fluid, allowing to define the {\em specific entropy} $s$ by the {\em local thermal equilibrium equation}:
\begin{equation}
\Theta ds =  d \epsilon + p d (1/r)  \, .        \label{re-termo0}
\end{equation}
The equations (\ref{ro-r-epsilon}), (\ref{c-masa}) and (\ref{re-termo0}) characterize the thermodynamic equilibrium of every volume element of the fluid,\footnote{This thermodynamic scheme for a relativistic perfect fluid is obtained as the adiabatic and Pascalian restriction of the Eckart's general local thermodynamic  equilibrium scheme \cite{Eckart} (see also \cite{Jou} for the present status of l.t.e.). Its causal character was first proved by A. Lichnerowicz \cite{Lichnero}.} and so, corresponding to a classical notion, define the {\em local thermal equilibrium} evolution (l.t.e.) of a perfect fluid.  We will also say that a perfect energy tensor evolves in l.t.e. if it verifies these equations.

If using (\ref{masa-energia}) we eliminate $\epsilon$ in
(\ref{re-termo0}), it is evident
that the integrability of $\Lambda$ is equivalent to the functional
dependence of the variables $\rho$, $p$ and $r$. Thus we have:
\begin{lemma}
A perfect energy tensor evolves in l.t.e. if, and only if, the equation (\ref{c-masa})
admits solutions $r(x) >0$ such that
\begin{equation}
dr \wedge d\rho \wedge dp = 0   \, .     \label{r-ro-p}
\end{equation}
Then, every pair $\{r,\Theta\}$, where $\Theta>0$ is an integrant divisor of
the one-form $(1/r) d \rho + (\rho+p) d(1/r)$, determines a
thermodynamic scheme. The specific entropy is then given, up to an
additive constant, by
\begin{equation}
\Theta ds = (1/r) d \rho + (\rho+p) d(1/r) \, .   \label{re-termo}
\end{equation}
\end{lemma}

From (\ref{r-ro-p}) and (\ref{re-termo}) we have that
$\rho$, $p$ and $s$ are dependent variables, and (\ref{con-eq2})
implies:
\begin{equation}
\dot{s} = -\frac{\rho+p}{\Theta r^2} \nabla\cdot(ru),       \label{r-s}
\end{equation}
equation that shows the intimate relation existing between the local
adiabatic evolution and matter conservation. With this, we have shown the
necessary condition of the following.
\begin{lemma}
A perfect energy tensor evolves in l.t.e. if, and only if, the equation
\begin{equation}
\dot{s} = 0             \label{spunto}
\end{equation}
admits solutions $s(x)$ such that
\begin{equation}
ds \wedge d\rho \wedge dp = 0  \, .   \label{s-ro-p}
\end{equation}
\end{lemma}

A way to show the sufficient condition is to consider a function $s(x)$
verifying equations (\ref{spunto}) and (\ref{s-ro-p}), and looking for a
function $r(x)$ solution of (\ref{c-masa}) and (\ref{r-ro-p}).

In the case of barotropic evolution, $d\rho \wedge dp = 0$, we can
take $r$ as an arbitrary solution of (\ref{c-masa}). In the case of non barotropic evolution, $d\rho \wedge dp \not= 0$, (\ref{s-ro-p})
implies that $s$ is a function of $\rho$ and $p$, $s=s(\rho,p)$, and then equation
(\ref{spunto}) can be written as:
\begin{equation}
s_{\rho}' \, \dot{\rho} + s_{p}' \, \dot{p} = 0  \, ,   \label{s-prima}
\end{equation}
where $s'_{\rho} = (s'_{\rho})_p = \big(\frac{\partial s}{\partial \rho}\big)_p$ and $s'_{p} = (s'_{p})_{\rho}$.

When the evolution is isoenergetic, $\dot{\rho}=0$, the second conservation equation (\ref{con-eq2})
implies $\theta=0$ and, then, every function $r(\rho)$ verifies 
(\ref{c-masa}) and (\ref{r-ro-p}).
When $\dot{\rho}\not=0$, from (\ref{s-prima}) we have:
\begin{equation}
\chi \equiv \frac{\dot{p}}{\dot{\rho}} = - \frac{s_{\rho}'}{s_{p}'} = \chi(\rho,p) \, . \label{chi}
\end{equation}
Then, based on (\ref{con-eq2}), we can write (\ref{c-masa}) in the form:
\begin{equation}
r_{\rho}'  + \chi(\rho,p) r_{p}' = \frac{r}{\rho+p} \, ,    \label{r-chi}
\end{equation}
which is an equation admitting solutions $r=r(\rho,p)$ that
fulfill (\ref{r-ro-p}). The sufficient condition of Lemma 2 is thus proven.

Making use of the above results, it is easy to give a
characterization of l.t.e. in terms of the sole hydrodynamic variables
$(u,\rho,p)$. Indeed, conditions $d\rho \wedge dp =0$ and $\dot{\rho}=0$ utilize exclusively
hydrodynamic variables and they assure the existence of associated
thermodynamic schemes without further restrictions.
In the generic case, when $d\rho \wedge dp \not=0$ and
$\dot{\rho}\not=0$, the
function $\chi \equiv \dot{p}/\dot{\rho}$ defined in (\ref{chi}) depends on $\rho$ and $p$. Then equation (\ref{spunto}) can be written as:
\begin{equation}
s_{\rho}'  + \chi(\rho,p) s_{p}' = 0   \, ,  \label{s-chi}
\end{equation}
which obviously admits solutions verifying (\ref{s-ro-p}).
All these cases may be linked in the following \cite{Coll-Ferrando-termo}.
\begin{theorem} \label{theo-LTE}
(Coll-Ferrando, 1989) A perfect energy tensor evolves in l.t.e. if, and only if,
\begin{equation}
(\dot{\rho} d \dot{p} - \dot{p} d \dot{\rho}) \wedge d\rho
\wedge dp = 0           \label{t1}
\end{equation}
\end{theorem}
Condition (\ref{t1}) identically holds in the {\em isoenergetic} case . Otherwise, when $\dot{\rho}\not=0$, the Theorem admits the following formulation: 
\begin{theorem} \label{theo-LTE-chi}
A non isoenergetic, $\dot{\rho}\not=0$, perfect energy tensor evolves in l.t.e. if, and only if, the space-time function $\chi \equiv \dot{p}/\dot{\rho}$, called {\em indicatrix of local thermal equilibrium}, depends only on the variables $p$ and $\rho$:  
\begin{equation}
d \chi \wedge d\rho \wedge dp = 0    \, .       \label{t2}
\end{equation}
\end{theorem}
As pointed out in \cite{cfERE}, this result has interesting physical consequences: 
\begin{itemize}
\item
This theorem says, in other words, that a relation of the form
\begin{equation} \label{causal}
\dot{p} = \chi (p,\rho )\dot{\rho}
\end{equation}
is a {\em minimal deterministic closure}\footnote{In the sense that the differential equations of the deterministic closure are of the lowest order.} to the equations $\nabla \cdot T = 0$ for any non barotropic perfect fluid.
\item
The "iff" character of Theorem \ref{theo-LTE-chi} implies that this
condition  constitutes an {\em alternative} definition of {\em  l.t.e.}.
And surprisingly enough, this alternative definition involves {\em only 
hydrodynamic, energetic} and {\em evolutive} concepts, but {\em not thermodynamic} ones. This means the important idea that the evolution in l.t.e. of a fluid generates the {\em emergence} of temperature and entropy: in spite of the usefulness of these functions, they are not {\em necessary} to verify the l.t.e. of the fluid.
\item
If the conditions of Theorem \ref{theo-LTE-chi} are verified, that is, if
the perfect fluid evolves in {\em  l.t.e.}, then, {\em and only then}, the indicatrix $\chi$ becomes a function of state, $\chi (\rho ,p)$, representing physically the square of the {\em speed of sound} $c_s$ in
the fluid,\footnote{Let us note that, because equation (\ref{causal}) is a minimal
deterministic closure, the relation (\ref{sound}) may be {\em directly} obtained by studying the propagation of infinitesimal perturbations by the system \{(\ref{con-eq1}),(\ref{con-eq2}),(\ref{causal})\}. From this point of view, the well known relation  $c_s^2= \left( {\partial p\over\partial\rho}\right)_s$ appears only {\em as  a constraint} for the {\em definition} of the entropy $s$.}
\begin{equation} \label{sound}
\chi (\rho ,p) \equiv  c^2_{s} \, .
\end{equation}

From the above interpretation, one has the following
{\em necessary condition of physical reality}:\footnote{This constraint can also be deduced from the {\em relativistic compressibility conditions} \cite{Israel} \cite{Lichnero-2}. Elsewhere \cite{C-F-S} we show that these conditions can be written in terms of the sole hydrodynamic variables by means of the indicatrix function $\chi(\rho,p)$.}
\begin{equation} \label{chi01}
0 \leq \chi \leq 1
\end{equation}
\end{itemize}


\section{Solving the inverse problem: admissible thermodynamic schemes}
\label{sec-invers-problem}

After the above comments, the inverse problem is tantamount to the analysis of how the speed of sound, given as a function of the hydrodynamic variables $(\rho, p)$, constraints the thermodynamic properties of the fluid. In other words: what perfect fluids evolve with a previously given indicatrix function $\chi(\rho,p)$?

We shall start with a perfect energy tensor $T\equiv(u,\rho,p)$ verifying (\ref{t1}) and look for the associated l.t.e. schemes. The richness and nature of these schemes will depend on the regularity of the given hydrodynamic data $(u,\rho,p)$. Consequently, several cases like barotropic or isobaroenergetic evolutions must be analyzed separately.


\subsection{Non barotropic perfect energy tensor: $d\rho \wedge dp \not=0$}
\label{subsec-nobar}

Under the non barotropic evolution assumption, we can consider the hydrodynamic variables $(\rho,p)$ as coordinates in the thermodynamic plane. 

When $(\rho,r)$ are independent variables, the form (\ref{re-termo}) of the l.t.e. equation shows that a thermodynamic
scheme is defined by a characteristic equation $s= s(\rho,r)$ that
determines all the other thermodynamic variables, so
\begin{equation}
p = - \rho - \frac{r(s_r')_{\rho}}{(s_{\rho}')_r} \equiv p(\rho,r) \; ,
\quad \quad \Theta = \frac{1}{r(s_{\rho}')_r} \equiv \Theta(\rho,r) \, .
\label{p-ro-r}
\end{equation}

In order to express the thermodynamic scheme in terms of the
hydrodynamic variables $(\rho,p)$, we can obtain $r(\rho, p)$ from the first equation in
(\ref{p-ro-r}), and putting it in the characteristic equation, we obtain
$s(\rho,p)$.
Conversely, these functions $r(\rho,p)$ and $s(\rho,p)$ determine a
thermodynamic scheme if we impose on them the form (\ref{re-termo}) of the l.t.e. equation. Using coordinates $(\rho,p)$, this last form is equivalent to:
\begin{equation}
r^2 \Theta s_p' = -r_p'(\rho+p) \, , \qquad
r^2 \Theta s_{\rho}' = r-r_{\rho}'(\rho+p) \, ,  \label{T-r-s-ro-p}
\end{equation}
which, under $\Theta \not= 0$, imply:
\begin{equation}
s_{\rho}'r_p' = s_p'\left[r_{\rho}'-\frac{r}{\rho+p}\right] \, .    \label{r-s-ro-p}
\end{equation}
Every pair $\{r(\rho,p),s(\rho,p)\}$ solution of (\ref{r-s-ro-p}) gives us a thermodynamic
scheme, with a temperature given by:
\begin{equation}
\Theta=\Theta^p(\rho,p) \equiv -\frac{r_p'}{r^2s_p'}(\rho+p), \quad {\rm
or} \quad \Theta=\Theta^{\rho}(\rho,p) \equiv
\frac{1}{s_{\rho}'}\left[\frac{1}{r}-\frac{r_{\rho}'}{r^2}(\rho+p)\right] ,
\label{t-ro-p}
\end{equation}
where only the first ({\em resp.} second) expression is valid in the case
$s_{\rho}'=0$ ({\em resp.} $s_p'=0$).

When $r=r(\rho)$, (\ref{re-termo}) gives $s=s(\rho)$ and, although
expressions (\ref{p-ro-r}) have no sense, (\ref{r-s-ro-p}) holds
and the second one in (\ref{t-ro-p}) remains valid.

Finally, from (\ref{masa-energia}), in both cases the specific internal energy is also
known in terms of $(\rho,p)$ variables:
\begin{equation}
\epsilon(\rho,p) = \frac{\rho}{r(\rho,p)} - 1  \, .      \label{e-ro-p}
\end{equation}
Thus, we have:
\begin{lemma}
For a non barotropic perfect energy tensor, the thermodynamic schemes are determined by a matter density
$r=r(\rho,p)$, solution of {\em (\ref{c-masa})}, and a specific entropy $s=s(\rho,p)$, solution of {\em (\ref{spunto})}, restricted by equation {\em (\ref{r-s-ro-p})}. Then, the temperature is given by {\em (\ref{t-ro-p})} and the specific internal energy by {\em (\ref{e-ro-p})}.
\end{lemma}
%


\subsubsection{Non barotropic and non isobaroenergetic evolution:
$d\rho \wedge dp \not=0$, $\; \dot{\rho}^2 + \dot{p}^2 \not = 0$}
\label{subsubsec-nobarA}

Let us consider a non barotropic perfect fluid energy tensor that
evolves in l.t.e. In order to obtain the admissible thermodynamic schemes
we must look for a solution $r(\rho,p)$ of the matter conservation equation (\ref{c-masa}) and  a
solution $s(\rho,p)$ of the entropy invariant evolution (\ref{spunto}) constrained by equation (\ref{r-s-ro-p}).

If the evolution is isonergetic, $\dot{\rho} = 0$, the second energy tensor conservation equation (\ref{con-eq2}) implies $\theta =0$, and then equations (\ref{c-masa}) and
(\ref{spunto}) write $\dot{p} r_p' = \dot{p} s_p'=0$. Thus, if in addition it is non isobaric, $\dot{p} \not= 0$, we have $r=r(\rho)$ and $s=s(\rho)$, and the equation (\ref{r-s-ro-p}) is obviously fulfilled.

If the evolution is non isoenergetic, $\dot{\rho} \not= 0$, the equation (\ref{t2}) of the
characterization Theorem expresses that the indicatrix,
$\chi \equiv \dot{p}/\dot{\rho}$, is a function of $\rho$ and
$p$: $\chi=\chi(\rho,p)$. Then the associated matter density $r(\rho,p)$
and the specific entropy $s(\rho,p)$ are submitted,
respectively,
to the first order partial differential equations (\ref{r-chi}) and
(\ref{s-chi}). Every pair $\{r,s\}$ solution to these equations verifies
(\ref{r-s-ro-p}). Thus, taking into account that (\ref{s-chi}) is the
homogeneous equation associated to (\ref{r-chi}), we can state:
\begin{proposition} \label{prop-nobar-noestatic}
Let $T$ be a non barotropic and non isobaroenergetic perfect energy tensor that
evolves in l.t.e.. The admissible thermodynamic schemes are defined by a matter
density $r(\rho,p)$ and a specific entropy $s(\rho,p)$ such that:

i) If $T$ is isoenergetic, $\dot{\rho} = 0$, they become arbitrary functions of $\rho$, $r=r(\rho)$ and $s=s(\rho)$.

ii) If $T$ is non isoenergetic, $\dot{\rho} \not= 0$, they are of the form $r=\bar{r}R(\bar{s})$ and $s=s(\bar{s})$,
where $\bar{r}(\rho,p)$ is any particular solution of equation {\em (\ref{r-chi})}, and
$R(\bar{s})$ and $s(\bar{s})$ are arbitrary functions of any particular
solution $\bar{s}(\rho,p)$ of equation {\em (\ref{s-chi})}, $\chi(\rho,p)$ being the
indicatrix function, $\chi \equiv \dot{p}/\dot{\rho}$.

For each thermodynamic scheme $\{r,s\}$ the temperature is given by
{\em (\ref{t-ro-p})} and the specific internal energy by {\em (\ref{e-ro-p})}.
\end{proposition}
Proposition \ref{prop-nobar-noestatic} fixes, in terms of arbitrary functions, the dimension of the set ${\bf F}_T$ of all perfect fluids $f$ in l.t.e. admitting a given non barotropic and non isobaroenergetic energy tensor $T$ as energetic evolution. Note that in both, isoenergetic and non isoenergetic cases, this dimension is controlled by two arbitrary functions of one real variable.

It is worth remarking that, in the isoenergetic case, $\dot{\rho} = 0$ and $\dot{p} \not=0$, the condition of physical reality (\ref{chi01}) does not hold because the speed of sound is infinite. Nevertheless, for the sake of formal completeness, we also take into account this degenerate case. 


\subsubsection{Non barotropic and isobaroenergetic evolution: $d\rho \wedge dp \not=0$, $\; \dot{\rho} = \dot{p} = 0$}
\label{subsubsec-nobarB}

According to (\ref{con-eq2}), along a non barotropic and isobaroenergetic evolution the expansion vanishes: $\theta=0$. This means that arbitrary functions $r=r(\rho,p)$ and $s=s(\rho,p)$ verify
equations (\ref{c-masa}) and (\ref{spunto}). Consequently, only equation
(\ref{r-s-ro-p}) must be imposed on them. One can take an arbitrary
$r=r(\rho,p)$ and look upon this equation as a first order partial
differential equation on $s=s(\rho,p)$:
\begin{equation}
s_{\rho}'  + \chi^r (\rho,p) s_{p}' = 0 \, , \qquad  \chi^r (\rho,p) \equiv \frac{1}{r_p'}\left[\frac{r}{\rho+p}- r_{\rho}' \right]  \, .  
\label{s-chi-r}
\end{equation}
Therefore, we can state:
\begin{proposition}
Let $T$ be a non barotropic and isobaroenergetic perfect energy tensor that
evolves in l.t.e.. The admissible thermodynamic schemes are defined by an
arbitrary matter density $r=r(\rho,p)$ and a specific entropy $s=s(\rho,p)$ such
that $s=s(\bar{s})$, $\bar{s}= \bar{s}(\rho,p)$ being any particular solution of {\em (\ref{s-chi-r})}.\\
For each thermodynamic scheme $\{r,s\}$ the temperature is given by
{\em (\ref{t-ro-p})} and the specific internal energy by {\em (\ref{e-ro-p})}.
\end{proposition}
In this case the richness of admissible l.t.e. schemes, i.e. the corresponding set ${\bf F}_{T}$, is tantamount to an arbitrary function of two variables and an arbitrary function of a sole variable.

Note that we can, alternatively, take an arbitrary specific entropy $s=s(\rho,p)$ and look for the matter density $r=r(\rho,p)$ solution of the equation (\ref{r-s-ro-p}):
\begin{equation}
r_{\rho}'  + \chi^s (\rho,p) r_{p}' = \frac{r}{\rho+p} \, , \qquad  \chi^s (\rho,p) \equiv -\frac{s_{\rho}'}{s_p'} \, .  \label{r-chi-s}
\end{equation}
Now we have an indeterminate indicatrix function $\chi = \dot{p}/\dot{\rho}$, but the state functions $\chi^r(\rho,p)$ and $\chi^s(\rho,p)$ give the square of the speed of sound. 

From the characteristic equation of an arbitrary perfect fluid one can calculate the state function $r(\rho, p)$ by using the l.t.e. equation (\ref{re-termo}), and equation (\ref{s-chi-r}) simply restricts the compatible specific entropies. Thus, we have: 
\begin{corollary} \label{cor-nonbar-iso}
Every non barotropic and isobaroenergetic perfect energy tensor $T$ represents the evolution in l.t.e. of any non barotropic perfect fluid.\footnote{Remember that the particularities of a perfect energy tensor are particularities of the {\em evolution} of the fluid described by the energy tensor but they are not necessarily particularities of its {\em material} constitution (see Section 1 and Figure 1).}
\end{corollary}
This Corollary shows that the set ${\bf F}_T$ of all perfect fluids $f$ admitting a non barotropic and isobaroenergetic $T$ as energetic evolution differs from the whole set ${\bf F}$ of all perfect fluids $f$ only by the set ${\bf F}_b$ of all barotropic fluids, ${\bf F}= {\bf F}_T \cup {\bf F}_b$.


\subsection{Barotropic perfect energy tensor: $d\rho \wedge dp =0$}
\label{subsec-bar}

When a barotropic perfect energy tensor $T$ has no constant energy density, $d\rho \not=0$, the barotropic condition can be stated as a {\em barotropic relation} of the form:
\begin{equation} \label{barotropic-rel}
p= \phi(\rho) \, .
\end{equation}

This barotropic relation can be interpreted as an equation of state of the medium represented by $T$, which holds independently of the considered particular evolution $T$. Then we say that this medium is an {\em (intrinsic) barotropic perfect fluid}. But it can also be interpreted as a particular evolution of non (intrinsic) barotropic media. For example, if a fluid with an equation of state $p= p(\rho, s)$ evolves at constant entropy, we have, for this particular evolution, $p = p(\rho,s_0) = \phi^s(\rho)$; or similarly, for a constant temperature evolution, we obtain $p=p(\rho, \Theta_0) = \phi^{\Theta}(\rho)$, where the superscripts $s$ and $\Theta$ recall the particular type of evolution.

Solving the inverse problem for a barotropic perfect energy tensor means determining both, the associated barotropic thermodynamic schemes, as well as the non barotropic schemes with the adequate barotropic evolution.

The barotropic perfect energy tensors are paradigmatic in the relativistic framework. They are obligatory in the conventional cosmology with the Friedmann-Lema\^itre-Robertson-Walker universes, and they offer the simplest and essential models for stellar interiors with static spherically symmetric perfect fluid solutions. Studying the inverse problem for these space-times from the above two barotropic points of view, will provide new physical interpretations of these solutions that could differ interestingly from those considered up to now.  


\subsubsection{Barotropic perfect fluids:  $p= \phi(\rho)$ as an equation of state}
\label{subsubsec-bar-intrinsec}

If we consider $\rho=\rho_0$ as an equation of state of the medium, the l.t.e.equation (\ref{re-termo}) leads to the following thermodynamic scheme: 
\begin{equation}
s=s(r), \qquad \Theta(r,p)= -\frac{\rho_0 + p}{r^2 s'(r)}, \qquad  \epsilon(r) = \frac{\rho_0}{r} -1, 
\end{equation}
$r$ being any solution of $\dot{r}=0$. This scheme has a doubtful physical meaning because it never satisfies the compressibility conditions (see \cite{C-F-S}).

Otherwise, if $d \rho \not=0$, we have a barotropic relation (\ref{barotropic-rel}), and then (\ref{re-termo}) leads to:
\begin{equation}
\Theta ds = \frac{\rho+\phi(\rho)}{r} \, d {\rm ln}
\left(\frac{G(\rho)}{r}\right),        \qquad  \quad
G(\rho) \equiv \exp \left[\int \! \!\frac{d \rho}{\rho+ \phi(\rho)}\right]  \, .  \label{re-termo-bar}
\end{equation}
Then, taking $(\rho, s)$ as coordinates in the thermodynamic plane, we obtain: 
\begin{equation}
r(\rho,s) = \frac{G(\rho)}{R(s)}, \quad 	 \Theta(\rho,s)= \frac{[\rho + \phi(\rho)]R'(s)}{G(\rho)},  \quad
 \epsilon(\rho,s) = \frac{\rho R(s)}{G(\rho)} -1  ,
\label{r-t-e-bar}
\end{equation}
where $R(s)$ is an arbitrary function. Thus, we have the answer to the inverse problem in $T$ restricted to the barotropic perfect fluids:
\begin{proposition} \label{prop-bar-in}
Let $T(u,\rho,p)$ be a barotropic perfect energy tensor with non constant energy, $p=\phi(\rho)$. The admissible barotropic thermodynamic schemes are defined by the characteristic equation $r(\rho,s) = G(\rho)/R(s)$, with $G(\rho)$ given in {\em (\ref{re-termo-bar})}, $s$ being an arbitrary solution of $\dot{s}=0$, and $R(s)$ an arbitrary real function. For each scheme $\{s,R(s)\}$, the matter density, the temperature and the specific internal energy are given in {\em (\ref{r-t-e-bar})}.
\end{proposition}
In this intrinsic barotropic case, the set ${\bf F}_T$ of all barotropic perfect fluids admitting $T$ as an energetic evolution is controlled by an arbitrary solution of $\dot{s}=0$ (an arbitrary function of three variables), and an arbitrary function of a sole variable. 

Note that the results in Proposition \ref{prop-bar-in} exclusively depend on the barotropic relation (\ref{barotropic-rel}). This means that a generic barotropic medium may or may not be in isobaroenergetic evolution. Nevertheless, if $s=s(\rho)$ is a state condition, we have one-dimensional thermodynamics, and necessarily isobaroenergetic evolution.

Several physically relevant barotropic media have been considered in the literature due to their applicability to relativistic astrophysics (see for example \cite{Anile} \cite{Rezzolla}). We study them briefly as particular cases of the generic barotropic media considered in Proposition \ref{prop-bar-in}.

We have, for example, {\em cold matter fluids}, which include the completely degenerate ideal Fermi gas. We can recover this case taking $\Theta=0$ in expressions (\ref{re-termo-bar}) and (\ref{r-t-e-bar}). Note that then $R(s)$ becomes a constant function and, consequently, we have:
\begin{corollary} \label{cor-bar-cold}
Any barotropic perfect energy tensor $T(u,\rho,p)$ with non constant energy, $p=\phi(\rho)$, represents the evolution in l.t.e. of a cold matter fluid ($\Theta=0$), with matter density given by:
\begin{equation}
r(\rho) = r_0 \exp \left[ \int \! \!\frac{d \rho}{\rho+ \phi(\rho)}\right]  \, .  \label{bar-cold}
\end{equation}
\end{corollary}
Now the set ${\bf F}_T$ of all cold matter fluids $f$ admitting $T$ as an energetic evolution is controlled by a sole constant $r_0$.

Another interesting example of barotopic perfect fluid is a gas of particles in thermal equilibrium with radiation when the particle energy density is negligible compared to the radiation energy density. In this case the energy density and the pressure depend on the temperature alone:
\begin{equation}
\rho = \rho(\Theta)  \, , \qquad \quad p=p(\Theta)  \, .  \label{ro-p-T}
\end{equation}
It is assumed that the energy density is an effective function of temperature. Then (\ref{ro-p-T}) leads to a barotropic relation of type (\ref{barotropic-rel}). Now $(\Theta'_s)_{\rho} = 0$, and then (\ref{r-t-e-bar}) implies $R'(s) =$ constant. Thus, we can state:

\begin{corollary} \label{cor-bar-ultra}
Any barotropic perfect energy tensor $T(u,\rho,p)$ with non constant energy, $p=\phi(\rho)$, represents the evolution of a gas in l.t.e with dominant radiation. Moreover, $G(\rho)$ being given by {\em (\ref{re-termo-bar})} and $r$ being an arbitrary solution of {\em (\ref{c-masa})}, the temperature, the specific entropy and the matter density are given, respectively, by:
\begin{equation}
\Theta(\rho)= \frac{[\rho + \phi(\rho)]}{G(\rho)},  \qquad s(\rho,r)   = \frac{\rho+p}{\Theta r} =  \frac{G(\rho)}{r} , \qquad 	  \epsilon(\rho,r) = \frac{\rho}{r} -1  .
\label{s-t-e-ultra}
\end{equation}
\end{corollary}
In this case the set ${\bf F}_T$ of all gases of particles $f$ in thermal equilibrium with radiation admitting $T$ as an energetic evolution is controlled by an arbitrary function of three variables.

A particular case of this last Corollary is when the particles are highly relativistic. Then, we have the so-called {\em radiation fluid}, which satisfies a barotropic relation in the form:
\begin{equation}
\rho = 3p \,  .
\label{radiacio-bar}
\end{equation}
Now $G(\rho)= C \rho^{3/4}$ and, from expressions (\ref{s-t-e-ultra}), we recover the following known result:
\begin{corollary} \label{cor-bar-radiacio}
A perfect energy tensor $T(u,\rho,p)$ with barotropic relation $\rho=3p$, represents the evolution of a radiation fluid. The energy density, the pressure and the entropy density depend on the temperature as:
\begin{equation}
\rho = a \Theta^4 \, , \qquad \quad  p = \frac13 a \Theta^4 \, , \qquad  \quad S=rs = \frac43 a \Theta^3 \,  .
\label{s-t-e-radiacio}
\end{equation}
\end{corollary}
Now the set ${\bf F}_T$ of all radiation fluids $f$ admitting $T$ as an energetic evolution is controlled by a sole constant $a$.\footnote{If the radiation fluid is a gas in l.t.e. with radiation, the constant $a$ is the Stefan-Boltzmann constant $a_R$. If it is a fluid of massless neutrinos $a= (7/16) a_R$ and, for a fluid of ultrarelativistic electron-positron pairs, $a= (7/8) a_R$.}


\subsubsection{The barotropic relation $p= \phi(\rho)$ as an evolution condition}
\label{subsubsec-bar-evolution}

Given a generic perfect fluid with characteristic equation $r=r(\rho, s)$, we can consider any barotropic relation $p= \phi (\rho)$ as an evolution condition. The l.t.e. equation (\ref{re-termo}) implies $s=c(\rho)$, this evolution constraint being defined by the relation:
\begin{equation} 
p(\rho, c(\rho)) =\phi (\rho)  \, , \qquad \quad   p(\rho, s) \equiv \frac{r(\rho, s)}{(r'_\rho)_s} - \rho  \,  .  \label{bar-constraint}
\end{equation}
Moreover, if $d \rho=0$, we can see $\rho = \rho_0$ as an evolution constraint. Consequently, a barotropic evolution does not restraint the thermodynamic scheme. 

Nevertheless, when $c'(\rho) \not=0$ we have $\dot{\rho}=0$, and then $\dot{p}=0$, and the evolution is necessarily isobaroenergetic. Thus:
\begin{proposition} \label{prop-bar-state}
Every barotropic and isobaroenergetic perfect energy tensor $T$ represents the evolution in l.t.e. of any perfect fluid.\footnote{See footnote 16.} 
\end{proposition}
This Proposition states that the set ${\bf F}_T$ of all perfect fluids $f$ admitting a barotropic and isobaroenergetic $T$ as energetic evolution is the full set of the perfect fluids, ${\bf F}_T={\bf F}$.

We can consider three restricted problems which are more interesting, from a practical point of view, than the generic result above:
\begin{description}
\item[i)]
A specific direct problem: to determine the barotropic and isobaroenergetic perfect energy tensors (namely, the barotropic relation $p=\phi(\rho)$) corresponding to the evolution of a given specific family of perfect fluids $r=r(\rho,s)$ evolving with a given constraint $s=c(\rho)$:
\begin{equation} 
\phi(\rho) = \frac{r(\rho, c(\rho))}{(r'_\rho)_s(\rho, c(\rho))} - \rho  \,  .  \label{bar-constraint-b}
\end{equation}
\item[ii)]
A specific inverse problem: to determine the perfect fluids (namely, their characteristic equation $r=r(\rho,s)$) for which a given barotropic and isobaroenergetic perfect energy tensor (namely, with a given barotropic relation $p=\phi(\rho)$) describes a prescribed constrained evolution $s=c(\rho)$. The characteristic equation $r=r(\rho,s)$ is subjected to the restriction:
\begin{equation} 
(R'_\rho)_s(\rho, c(\rho)) = \frac{1}{\rho+\phi(\rho)} \, , \quad R(\rho,s) = \ln r(\rho,s)  \,  .  \label{bar-constraint-c}
\end{equation}
\item[iii)] 
A problem of evolution: given a specific family of perfect fluids $r=r(\rho,s)$ and a particular barotropic relation $p=\phi(\rho)$, to obtain the condition $s=c(\rho)$ which constrains the evolution. This amounts to obtain a solution to (\ref{bar-constraint}).
\end{description}
In the direct and inverse specific problems i) and ii) it will be worth considering evolution constraints with a remarkable physical meaning, like evolutions at constant temperature or at constant entropy. 

On the other hand, when $c'(\rho)=0$, that is, in an evolution at constant entropy $s=s_0$, the evolution is not, necessarily, isoenergetic and, for a specific family of perfect fluids with characteristic equation $r=r(\rho,s)$, we have a barotropic evolucion with barotropic relation:
\begin{equation}  \label{s=s0}
\phi(\rho) = \frac{r(\rho, s_0)}{(r'_\rho)_s(\rho, s_0)} - \rho  \,  .  \label{bar-constraint-entropy}
\end{equation}
Moreover, this condition constraints the thermodynanic schemes $r=r(\rho,s)$ if we impose a barotropic relation $p=\phi(\rho)$. Then, we have:
\begin{proposition} \label{prop-bar-state-entropy}
A barotropic ($p=\phi(\rho)$) and non isoenergetic ($\dot{\rho}\not=0$) perfect energy tensor $T$ represents the evolution in l.t.e. of the perfect fluids with characteristic equation $r=r(\rho,s)$ restricted by:
\begin{equation} 
(R'_\rho)_s(\rho, s_0) = \frac{1}{\rho+\phi(\rho)} \, , \quad R(\rho,s) = \ln r(\rho,s)  \,  .  \label{bar-constraint-c}
\end{equation}

\end{proposition}

\subsection{A summary of the inverse problem}
\label{subsec-table}

In the table below we summarize the results on the inverse problem presented in this section. Note that our study provides a classification of the perfect energy tensors ${\bf T}_{\bf F}$ in four classes, which have different solutions to the inverse problem.

The first rows at the top of the table present the conditions on the hydrodynamic variables $\{u,\rho,p\}$ defining these four classes, that is, four disjoint subsets of $T$. The following row contains the equations that we must necessarily solve in order to obtain the associated thermodynamic schemes. The row below shows the richness of these thermodynamics, that is the dimension of the set ${\bf F}_T$, and gives the expression for matter density and specific entropy. The last row presents the expression of temperature and specific internal energy.


\begin{longtable}{|c|c|c|c|c|}\hline
\multicolumn{5}{|c|} {$\begin{array}{c}  \\[-3mm]   \qquad  ( \dot{\rho}\,
\dif \dot{p} - \dot{p} \, \dif \dot{\rho} ) \wedge \dif \rho \wedge
\dif p =0
\qquad  \end{array} $}  \\[2mm] \hline
 {\multirow{3}{*}{ {\scriptsize $\begin{array}{c} \\[-1.6mm] {\rm C} \\[-0.5mm] {\rm L} \\[-0.5mm] {\rm A} \\[-0.5mm] {\rm S}  \\[-0.5mm] {\rm S}\\[-0.5mm] {\rm E} \\[-0.5mm] {\rm S} \end{array} $} }} &\multicolumn{3}{c|}  {$\begin{array}{c} \\[-3mm] \quad \dif \rho \wedge \dif p \neq 0 \quad \end{array} $ } &
  {\multirow{3}{2cm}{ $\dif \rho \wedge \dif p = 0 $ \\[2mm] $\quad p
\equiv \phi(\rho)$   } }                 \\[2mm]\cline{2-4}
                         &  {\multirow{2}{*} {$\begin{array}{c} \\ \quad \dot{\rho} \neq 0 \end{array}$}  } &\multicolumn{2}{c|} {$\begin{array}{c} \\[-3mm] \quad \dot{\rho} =0 \quad \\[1mm] \end{array} $}    &
                         \\ \cline{3-4}
                         &  & {$\begin{array}{c} \\[-2mm] \dot{p} =0 \\[1mm] \end{array}$}   & {$\begin{array}{c} \\[-2mm]  \dot{p} \neq 0 \\[1mm] \end{array}$} &  \\[1mm]\hline
{ {\scriptsize $\begin{array}{c} {\rm E} \\[-0.4mm] {\rm Q} \\  {\rm U} \\[-0.4mm] {\rm A} \\[-0.4mm] {\rm T}  \\[-0.4mm] {\rm I}  \\[-0.4mm] {\rm O}  \\[-0.4mm] {\rm N} \\[-0.4mm] {\rm S} \end{array} $} }  &
 \hspace*{-1.3mm}$\begin{array}{c} \\[-2mm]\chi \! \equiv \!  \frac{\dot{p}}{\dot{\rho}} \! = \!
\chi(\rho, p ) \\[3mm] \bar{r}(\rho, p)  ,   \  \bar{s}(\rho, p)\!: 
\\[3mm]
\bar{r}^{\prime}_{\rho} \! +\!  \chi \, \bar{r}^{\prime}_{p} \!
= \! \frac{\bar{r}}{\rho + p}
\\[3mm]
\bar{s}^{\prime}_{\rho}  + \chi \, \bar{s}^{\prime}_{p} =0
\\[3mm]
  \end{array}$  & \hspace{-2mm} $\begin{array}{c}  s = s(\rho, p)   \\[3mm]   \bar{r}(\rho, p): \\[3mm]   {\scriptsize \! s^{\prime}_{\rho} {\bar{r}}^{\prime}_p \! = \!
 s^{\prime}_{p} \Big[\! {\bar{r}}^{\prime}_{\rho} \! - \! \frac{\bar{r}}{\rho + p} \! \Big]\! \! \!} \\  \end{array} $
  & none & $ \begin{array}{c}   {\scriptsize G(\rho) \! = \! \exp \! \Big[\!{\int \!\! \frac{\dif \rho}{\rho +
  \phi(\rho)}}} \Big]
\\[4mm]
s_i , \quad  i= 1, 2 , 3 : \\[2mm]  \dot{s}_{i}  =0   \\     \end{array}
 $  \\ \hline
 {\multirow{2}{2mm}{\scriptsize $\begin{array}{c}\\[-9mm] {\rm T}  \\[-0.4mm] {\rm E}   \\[-0.4mm] {\rm R}   \\[-0.4mm] {\rm M} \\[-0.4mm] {\rm O}  \\[-0.4mm] {\rm D}  \\[-0.4mm]  {\rm I}  \\[-0.4mm] {\rm N} \\[-0.4mm]
 {\rm A} \\[-0.4mm] {\rm M}  \\[-0.4mm] {\rm I}   \\[-0.4mm] {\rm C}    \\[-0.4mm]  {\rm S}    \\[-0.4mm]  \end{array}$ } } &  $\begin{array}{c} \\[-2mm] \{ r, s \} \\[2mm] r = \bar{r} \ R(\bar{s}) \\[2mm] s= s(\bar{s})\\[1mm]  \end{array} $
 &   $\begin{array}{c} \\[-2mm]  \{ r, s \} \\[2mm] s = s(\rho, p)  \\[2mm]  r=  \bar{r}\  R(s) \\[1mm]  \end{array} $
  & $\begin{array}{c} \\[-2mm] \{ r, s \} \\[2mm] \! \! s \! = \! s(\rho ) \! \\[2mm] \! \! r \!=  \! r(\rho)\! \! \\[1mm] \end{array} $   &
  $\begin{array}{c} \\[-2mm] \{ r, s \} \\[2mm]  r = \frac{G(s)}{R(s)} \\[2mm]  s=  s(s_1 , s_2, s_3 ) \\[1mm]  \end{array} $ \\ \cline{2-5}
                           & \multicolumn{3}{c|} {\footnotesize {$\begin{array}{c} \\[-2mm] \! \Theta(\rho,p) \! = \!  \frac{r^{\prime}_p}{r^2 s^{\prime}_p} (\rho + p)   \ \
                           {\rm or} \ \ \Theta(\rho, p) \!=\! \frac{1}{s^{\prime}_{\rho}} \! \left[ \frac{1}{r} \!-\!  \frac{r^{\prime}_\rho}{r^2} (\rho + p) \right] \!  \\[3mm]
                           \epsilon(\rho, p) = \frac{\rho}{r(\rho, p)} - 1 \\[2mm]  \end{array}$  } }&
                            {\footnotesize $\begin{array}{c} \\[-2mm] \Theta(\rho, s)\!=\! \frac{[\rho + \phi(\rho)] R^{\prime} (s)}{G(\rho)} \\[3mm]
          \epsilon(\rho, s)= \frac{\rho R(s)}{G(\rho)} - 1  \\[2mm]  \end{array} $}
\\ \hline

\end{longtable}


\section{When is a perfect energy tensor the evolution of an ideal gas?}
\label{sec-ideal-gas}

Theorems \ref{theo-LTE} and \ref{theo-LTE-chi} allow us to know if a perfect energy tensor models the evolution of some perfect fluids, but they do not offer information about the specific physical properties of such fluids. If we are interested in a particular family of fluids ${\bf G}$, we must solve the corresponding specific direct and inverse problems: i) to obtain a deductive criterion to detect if a given perfect energy tensor $T$ performs the evolution of a perfect fluid of this family, namely, to determine ${\bf T}_{\bf G}$, and ii) to obtain all the perfect fluids of this family for which $T$, fulfilling this criterion, follows a particular evolution, namely, to determine ${\bf G}_T$, for $T \in {\bf T}_{\bf G}$.

Here, to show how one can solve these problems, we consider the paradigmatic family ${\bf G}$ of ideal gases. A (generic) ideal gas is characterized by the equation of state:
\begin{equation}
p = kr\Theta  \, , \qquad \quad    k \equiv {k_B \over m} \,  .  \label{gas-ideal}
\end{equation}
If we take $(\rho,r)$ as coordinates in the thermodynamic plane, the form (\ref{re-termo}) of the l.t.e. equation leads to the following ideal gas characteristic equation:\footnote{The independence of the variables $(\rho,r)$ is a necessary requisite to avoid degenerate one-dimensional thermodynamics.}
\begin{equation}
s = s(\rho,r) = \xi(e) - k \ln r \, , \qquad  e \equiv \frac{\rho}{r} = 1 + \epsilon  \,  ,  \label{gas-ideal-car-eq}
\end{equation}
where $\xi(e)$ is an effective function of the {\em specific energy} $e$. Moreover, the temperature depends on $e$ as:\footnote{When $\Theta'(e) \not=0$ we have $e=e(\Theta)$, and then the internal energy density is a function of the temperature: $\epsilon = \epsilon(\Theta) = e(\Theta) -1$. Usually, this property is supposedly satisfied by an ideal gas. Nevertheless, our formal study also includes the case $\Theta = \Theta_0$ as a permissible equation of state.}
\begin{equation}
\Theta \equiv \Theta(e) = \frac{1}{\xi'(e)}   \,  .  \label{gas-ideal-T-e}
\end{equation}
Then, from (\ref{gas-ideal}), (\ref{gas-ideal-car-eq}) and (\ref{gas-ideal-T-e}), we obtain that the hydrodynamic variable $\pi=p/\rho$ is also a function of the specific energy $e$:
\begin{equation}
\pi = \pi(e) \equiv {k\Theta(e) \over e}  , \qquad \qquad \pi \equiv \frac{p}{\rho}  \, .
\label{pi-e}
\end{equation}
%


\subsection{Barotropic ideal gases}
\label{subsec-barideal}

From equation (\ref{pi-e}) we have that the function $\pi(e)$ is constant for an ideal gas if,  and only if, $e(\Theta)=c_v \Theta$, where $c_v$ is a constant. Then (\ref{pi-e}) becomes:
\begin{equation}
\frac{p}{\rho} \equiv \pi  \equiv \pi(e) = {k \over c_v} \equiv \gamma-1  \, ,  \label{pi=C}
\end{equation}
and we obtain that the ideal gas is a barotropic media with a {\em relativistic $\gamma$-law}, $p= (\gamma-1) \rho$, as a barotropic equation of state. 

Are there other barotropic ideal gases? If $p=\phi(\rho)$ is an equation of state for an ideal gas, the barotropic scheme (\ref{re-termo-bar}), (\ref{r-t-e-bar}), and the ideal gas relations (\ref{gas-ideal}), (\ref{gas-ideal-car-eq}) must be compatibles. Then, necessarily, we obtain that $e= c_v \Theta$ and $p= (\gamma-1) \rho$.  Consequently, we extend a known result \cite{coll}:
\begin{proposition}
The unique barotropic ideal gases are those that verify $\epsilon(\Theta) =
c_v \Theta-1$. Then the barotropic equation of state is a $\gamma$-law
$p=(\gamma-1)\rho$.
\end{proposition}
%


\subsection{Non barotropic evolution of an ideal gas}
\label{subsec-nobarideal}

Now $(\rho,p)$ are coordinates in the thermodynamic plane. In addition, after the study above on barotropic ideal gases, in the non barotropic case we have $\pi'(e)\not=0$, and thus we can determine the inverse function $e=e(\pi)$. Then, taking into account the l.t.e.equation (\ref{re-termo}), we can use this function in order to write all the thermodynamic variables in terms of the hydrodynamic ones $(\rho,p)$. In particular, the speed of the
sound can be determined by using $v_s^2(\rho,p)= -{s_{\rho}' \over s_p'}$. We summarize these expressions in the following.
\begin{lemma}  \label{lemma-ideal}
Consider a non barotropic ideal gas characterized by the characteristic equation {\em (\ref{gas-ideal-car-eq})}, and the temperature $\Theta=\Theta(e)$ {\em (\ref{gas-ideal-T-e})}, with $e \not= c_v \Theta$. Let $e=e(\pi)$ be the inverse function of the $\pi(e)$ given in {\em (\ref{pi-e})}. In terms of the hydrodynamic variables $(\rho,p)$, the specific internal energy
$\epsilon$, the temperature $\Theta$, the matter density $r$, the specific entropy $s$ and the speed of the sound $c_s$ are given, respectively, by:
\begin{eqnarray}
\epsilon(\rho,p) = \epsilon(\pi) \equiv e(\pi)-1 \, , \qquad \quad
\Theta(\rho,p) = \Theta(\pi) \equiv {\pi \over k} e(\pi) \, , \label{e-t-ideal} \\
r(\rho,p) = {\rho \over e(\pi)} \, ,  \qquad  \qquad    \qquad  \
\quad s(\rho,p) = k \ln \frac{f(\pi)}{\rho} \, , \quad  \quad \ \  \label{r-s-ideal} \\
c_s^2(\rho,p) = \pi + {1 \over \phi(\pi)} \equiv \bar{\chi}(\pi) \not= \pi , \qquad \qquad \qquad 
\label{vel-so} 
\end{eqnarray}
where
\begin{equation}
f(\pi) \equiv f_0 \exp\{ \! \int \!\! \phi(\pi)d\pi\} \, , \qquad \quad  \phi(\pi) \equiv \frac{(\pi + 1) e'(\pi)}{\pi e(\pi)} .  \quad   \label{f-phi}
\end{equation}
\end{lemma}

The general study of the non barotropic case presented in Sect. \ref{subsec-nobar}, and Lemma \ref{lemma-ideal} above for the ideal gas, imply that an isoenergetic ($\dot{\rho}=0$) evolution is, necessarily, a isobaroenergetic one, ($\dot{p}=0$). Then, the only restrictions on the thermodynamic scheme are given by the expressions in Lemma above, $e(\pi)$ being an arbitrary function. Thus:
\begin{proposition}
The necessary and sufficient condition for a non barotropic and
isoenergetic ($\dot{\rho} =0$) perfect energy tensor $T=(u,\rho,p)$ to represent the l.t.e. evolution of an
ideal gas is to be isobaroenergetic: $\dot{\rho}=0$, $\dot{p}=0$.
Then $T$ represents the evolution in l.t.e. of any non barotropic
ideal gas, and the admissible thermodynamic schemes are defined by the specific internal energy
$\epsilon$, the temperature $\Theta$, the matter density $r$, the specific entropy $s$ and the speed of the sound $c_s$ given in {\em (\ref{e-t-ideal})}, {\em (\ref{r-s-ideal})}, {\em (\ref{vel-so})} and {\em (\ref{f-phi})}, $e(\pi)$ being
an arbitrary effective function of $\pi=p / \rho$.
\end{proposition}
Note that to each specific ideal gas, determined by the function $\xi(e)$ in (\ref{gas-ideal-car-eq}) corresponds a particular function $e(\pi)$, related through relations (\ref{gas-ideal-T-e}) and (\ref{pi-e}). And conversely, each $e(\pi)$ generates a specific ideal gas scheme. Thus, accordingly with Corollary \ref{cor-nonbar-iso},  this Proposition shows that the set ${\bf G}_T$ of all ideal gases $f$ admitting a non barotropic and isobaroenergetic $T$ as energetic evolution differs from the whole set ${\bf G}$ of all ideal gases $f$ only by the set ${\bf G}_b$ of all barotropic ideal gases, ${\bf G}= {\bf G}_T \cup {\bf G}_b$.

When $\dot{\rho} \not=0$, (\ref{vel-so}) means that the indicatrix function $\chi=\dot{p}/\dot{\rho}$, which coincides with the square of the speed of sound, must be a function of $\pi$. Conversely, if $\chi=\chi(\pi)$, the function $e(\pi)$ is constrained by (\ref{vel-so}), (\ref{f-phi}). Taking into account these considerations we can state:
\begin{theorem} \label{theo-nobar-idealgas}
The necessary and sufficient condition for a non barotropic and non
isoenergetic ($\dot{\rho} \not =0$) perfect energy tensor $T=(u,\rho,p)$ to represent the l.t.e. evolution of an
ideal gas is that the indicatrix function $\chi\equiv \dot{p}/\dot{\rho}$ be a function of the variable $\pi \equiv p/\rho$,
$\chi=\chi(\pi) \not= \pi$:
\begin{equation}
d \chi \wedge d\pi = 0 , \qquad \chi \not= \pi  \, . \label{chi-pi}
\end{equation}
\end{theorem}
This Theorem solves the specific direct problem for the non barotropic ideal gases  $\bar{\bf G}$, that is, it characterizes the set ${\bf T}_{\bar{\bf G}}$.

\begin{proposition} \label{prop-nobar-idealgas}
A non barotropic and non isoenergetic perfect energy tensor $T=(u,\rho,p)$ verifying {\em (\ref{chi-pi})} ($\chi=\chi(\pi)$) represents
the l.t.e. evolution of the ideal gas with specific internal energy $\epsilon$, temperature $\Theta$, matter density $r$, and specific entropy {\em (\ref{e-t-ideal})} and {\em (\ref{r-s-ideal})}, the generating functions $e(\pi)$ and $f(\pi)$ being, respectively,
\begin{equation}
e(\pi) = e_0 \exp\{\! \! \int \! \! \psi(\pi)d\pi \} \, , \qquad  \qquad \psi
(\pi) \equiv \frac{\pi}{(\chi(\pi)-\pi)(\pi+1)} \, \ , \label{e-pi} 
\end{equation}
where
\begin{equation}
f(\pi) = f_0 \exp\{\! \! \int \! \! \phi(\pi)d\pi\} \, , \qquad \qquad
\phi(\pi) \equiv {1 \over \chi(\pi)-\pi} \, . \qquad \qquad  \
\label{f-pi}
\end{equation}
\end{proposition}
This Proposition solves the specific inverse problem by determining ${\bar{\bf G}}_T$ for a non barotropic and non isoenergetic energy tensor $T$. Note that a three parameter family $(f_0, e_0, k)$ of ideal gases can be associated with a perfect energy tensor with an indicatrix function $\chi$ (square of the speed of sound) subjected to the constraint (\ref{chi-pi}). The first one, $f_0$, fixes the origin of entropy, and we can consider that the different values correspond to a sole ideal gas.
The second parameter, $e_0$, modifies the specific energy in a constant factor and, consequently, the temperature and the {\em specific volume} $1/r$ change in the same factor. Be aware that $e_0$ settles the origin of internal energy, which change as $\epsilon \rightarrow e_0 \epsilon + e_0 -1$. Finally, the third one, $k=k_B/m$ determines, for fixed $e_0$, the mass of gas particles. Note that the hydrodynamic   variable $\pi$ fixes the product $k\Theta$; thus, changing the temperature in a factor, and the entropy in the inverse factor, we can adjust any value of the mass particles.


\subsection{The extended inverse problem for an ideal gas indicatrix $\chi(\pi)$}
\label{subsec-inverse-idealgas}

Theorem \ref{theo-nobar-idealgas} solves a specific direct problem: it characterizes the perfect energy tensors $T$ that are particular evolutions of a (non barotropic and non isoenergetic) ideal gas. And Proposition \ref{prop-nobar-idealgas} solves the associated specific inverse problem: it provides the ideal gas schemes associated with one of these $T$. This last result is useful when we are interested in ideal gases. Nevertheless, it does not solve the general inverse problem for $T$ satisfying constraint (\ref{chi-pi}): what perfect fluids ${\bf F}_T$, in addition to ideal gases, evolve with an ideal gas indicatrix $\chi(\pi)$? The answer to this extended inverse problem is given by Proposition \ref{prop-nobar-noestatic}: we must find particular solutions $\bar{r}$ and $\bar{s}$ to equations (\ref{r-chi}) and (\ref{s-chi}), respectively, with $\chi=\chi(\pi)$. But these particular solutions are provided by the results on ideal gases above. Thus, from Propositions \ref{prop-nobar-idealgas} and \ref{prop-nobar-noestatic} we get:
\begin{corollary} \label{cor-invers-chi-ideal}
Let $T=(u,\rho,p)$ be a non barotropic and non isoenergetic perfect energy tensor that
satisfies {\em (\ref{chi-pi})}, $\chi=\chi(\pi)$. The admissible thermodynamic schemes are defined
by a matter density $r=\bar{r}R(\bar{s})$ and a specific entropy $s=s(\bar{s})$, where $\bar{r}(\rho,p)$ and $\bar{s}(\rho,p)$ are given in {\em (\ref{e-t-ideal})} and {\em (\ref{r-s-ideal})}, $e(\pi)$ and $f(\pi)$ depending on $\pi$ as {\em (\ref{e-pi})} and {\em (\ref{f-pi})}, and
$R(\bar{s})$ and $s(\bar{s})$ being arbitrary functions.

For each thermodynamic scheme $\{r,s\}$ the temperature is given by (\ref{t-ro-p}) and the specific internal energy by {\em (\ref{e-ro-p})}.
\end{corollary}
%


\subsection{Barotropic evolution of a non barotropic ideal gas}
\label{subsec-barevol-idealgas}

In Sect. \ref{subsubsec-bar-evolution} we have stated that every barotropic and isobaroenergetic perfect energy tensor represents a possible evolution of any perfect fluid, and consequently, of any ideal gas. This barotropic evolution, $p=\phi(\rho)$, results from a constraint that, generically, may have an unclear interpretation. It is more interesting to impose a physically relevant constraint and to analyze the restricted direct and inverse problems stated in the points i) and ii) that follow Proposition \ref{prop-bar-state}.

As an example, let us consider the evolution of an ideal gas at the constant temperature $\Theta_0$. Then, from the second relation in (\ref{e-t-ideal}) we obtain:
\begin{equation}  \label{t-constant}
\pi e(\pi) = k \Theta_0 \, ,
\end{equation}
Then, necessarily, $p = \pi_0 \rho$, where $\pi_0$ is any solution to equation (\ref{t-constant}). And conversely, the barotropic evolution $p = \pi_0 \rho$ represents the isobaroenergetic evolution at a constant temperature of any ideal gas. If we add a specific energy density $e(\pi)$, the ideal gas scheme is determined by relations (\ref{e-t-ideal}) and (\ref{r-s-ideal}). Thus we have:
\begin{proposition} \label{prop-t-constant}
A perfect energy tensor $T=(u,\rho,p)$ represents the evolution at constant temperature of an ideal gas if, and only if, it is isobaroenergetic, $\dot{\rho} = \dot{p}=0$, and the following barotropic relation holds:
\begin{equation}
p = \pi_0 \rho \, .
\label{t-constant-p-rho}
\end{equation}
Conversely, the barotropic evolution $p = \pi_0 \rho$ represents the isobaroenergetic evolution at constant temperature of any ideal gas. For a given specific energy density $e(\pi)$, the product $k \Theta_0$ is constrained by the condition $\pi_0 e(\pi_0) = k \Theta_0$, and the specific internal energy $\epsilon$, the matter density $r$, and the specific entropy $s$ are given in {\em (\ref{e-t-ideal})} and {\em (\ref{r-s-ideal})}.
\end{proposition}
%


\section{Remarks and applications.}
\label{sec-remarks}

Problems in theoretical physics, as well as in mathematics, may be solved in many different, non equivalent, ways. Think, for example, on the conditions for a metric to be flat, and consider the three classical answers: "when and
only when there exists a coordinate system in which the components of the metric tensor are constant", "when and only when the metric is invariant by the corresponding \break (pseudo-)Euclidean group" and
"when and only when its Riemann tensor vanishes". In spite of the "when and only when" bijective correspondence of the answers to the same problem, the three answers are not in fact equivalent, because the background set of mathematical elements needed for each of them is different. For example, the first answer is {\em non-covariant} (coordinate-dependent) and {\em non-deductive} (cannot be generically checked deductively), meanwhile the second answer, although may be checked covariantly and deductively, is {\em non-intrinsic} (needs an element, the group, not explicit in the setting of the problem and not deductively attached to it). Only the third answer is simultaneously related to the data of the problem (and only to them) in an {\em intrinsic} (and consequently {\em covariant}), {\em deductive}, {\em explicit} (the expression of the curvature tensor in terms of the data, i.e. the metric, is explicitly known) and {\em algorithmic} (the curvature tensor is algorithmically related to the metric) way.

Of course, in relativity one can find problems solved in intrinsic (and thus covariant, if so were the statement of the problems) and deductive ways, but we want here to quote two paradigmatic papers on attractive problems which try to solve them deliberately in an intrinsic and covariant way. The first one is the work by Rainich \cite{Rainich} on the non null electromagnetic field (see Sect. \ref{subsec-Rainich} below). The second one is a note by Takeno \cite{Takeno-1} (see also \cite{Takeno-2}) where, in Takeno's words,"...a theory concerning the
discrimination of the spherically symmetric spacetimes has been constructed. Although
it is not of the ideal form". The shortcoming to achieve this "ideal form" comes from the unknowledge of explicit expression for the metric invariants of the curvature tensor used in the intrinsic characterization. This lack has been overcoming in a paper \cite{fs-SSST} where the interest of solving problems in an IDEAL form has been outlined. The use of the appellation IDEAL (as an acronym) seems to be adequate when the conditions obtained are Intrinsic, Deductive (no inference process is necessary), Explicit and ALgorithmic (a flow chart with a finite number of steps can be built).

The answers presented in this paper to the direct and inverse problems stated in the introduction have these characteristics of IDEAL solutions. They are intrinsic (i.e. involve only the data of the sole perfect energy tensor T, or those of the proper hydrodynamic and thermodynamic quantities defining the thermodynamic fluid) and thus covariant (i.e. involve T as a tensor, or the proper quantities as scalars or the unit vector velocity, in coordinate-free form), deductive (i.e. they do not need any inductive process to be verified), explicit  (i.e. they may be verified by direct substitution of the data and of deductive differential concomitants of them) and algorithmic (i.e. involve a finite number of steps for their verification). These evident conceptual and practical qualities allow us to apply our results in diverse contexts. Now we comment on some of these applications and future prospects.


\subsection{Rainich-like theories for perfect fluid solutions}
\label{subsec-Rainich}

The Rainich work \cite{Rainich} on the non null electromagnetic field provides, among others, three interesting problems: (i) to express Maxwell equations not in terms of field variables but in terms of the energy variables, (ii) to obtain the algebraic conditions
and the additional differential restrictions for a conserved symmetric tensor to be the
energy tensor of a Maxwell field, and (iii) to write all these conditions, via Einstein
equations, for the Ricci tensor considered as a metric concomitant. It is worth remarking that the Rainich approach proposes IDEAL solutions to these three problems.

A similar approach for the perfect fluid was developed in \cite{Coll-Ferrando-termo} by using two previous results. On the one hand, the characterization theorem of local thermal equilibrium, which we presented in the same paper \cite{Coll-Ferrando-termo}, and here we state as Theorem \ref{theo-LTE}.\break On the other hand, the complete algebraic study of a perfect energy tensor, which implies, not only its intrinsic characterization and the obtention of the eigenvalues, but also the covariant determination of the eigenvector associated to the simple eigenvalue \cite{bcm}. 

Note that solving restricted direct and inverse problems for a specific set of fluids is the first step for a Rainich-like theory for this set of fluids. For example, the results in Sect. \ref{sec-ideal-gas} will allow us to easily perform a Rainich approach for the ideal gas solutions of Einstein equations \cite{C-F-S}.


\subsection{Physical meaning of known perfect fluid solutions. The ideal gas Stephani universes}
\label{subsec-Stephani}

Most of the perfect fluid solutions of Einstein equations have been obtained by considering adapted coordinates to the fluid velocity, by imposing symmetries or by assuming a type Petrov-Bel for the Weyl tensor. The algebraic requirements on the Ricci tensor that the field equations impose are sometimes supplemented with the energy conditions. Nevertheless, the physical meaning of most of the solutions remains unclear. Our results provide a method to test the physical reality of these solutions and to understand their thermodynamic properties. Indeed, our answer to the direct and inverse problems offers a complete algorithm, in four steps, to discern which metrics ${\bf S}_{\bf G}$ of a given family ${\bf S}$ of solutions of Einstein equations represent the evolution in l.t.e. of a specific set of fluids ${\bf G}$. For example, when ${\bf G}$ is the set of ideal gases we have the following steps:
\begin{description}
\item[Step 1] To calculate the coordinate dependence of the space-time functions $\pi \equiv p/\rho$ and $\chi\equiv \dot{p}/\dot{\rho}$ for the family of solutions ${\bf S}$. 
\item[Step 2] To determine the ideal gas subset ${\bf S}_{\bf G}$ of ${\bf S}$ by imposing the ideal gas hydrodynamic
condition (\ref{chi-pi}), $d \chi \wedge d \pi = 0$.
\item[Step 3] To obtain, in this subset, the explicit expression of the indicatrix function: $\chi = \chi(\pi)$.
\item[Step 4] To calculate, from $\chi = \chi(\pi)$, the generating functions $e = e(\pi)$ and $f = f(\pi)$ given in (\ref{e-pi})
and (\ref{f-pi}), and to obtain thereof the thermodynamic variables by using (\ref{e-t-ideal}) and (\ref{r-s-ideal}).
\end{description}
In \cite{C-F} we have used this algorithm to obtain the Stephani universes that can be interpreted as an ideal gas evolving in l.t.e. We have found that five classes of thermodynamic schemes are admissible, which give rise to five classes of regular models and three classes of singular models. 

Of course, for a different set of fluids the four steps in the above algorithm must be adapted taking into account the hydrodynamic characterization of such fluids. For the full set of perfect fluids ${\bf F}$ we must use the generic characterization presented in Theorem \ref{theo-LTE}.



\begin{acknowledgements}
This work has been supported by the Spanish ``Ministerio de
Econom\'{\i}a y Competitividad", MICINN-FEDER project FIS2015-64552-P.
\end{acknowledgements}

\end{document}